# Negative magnetic relaxation in superconductors

E.P.Krasnoperov

NRC Kurchatov Institute, 123182, Moscow
MPTI, 141700, Russia, Moscow reg.

*It was observed that the trapped magnetic moment of HTS tablets or annuli increases in time (negative relaxation) if they are not completely magnetized by a pulsed magnetic field. It is shown, in the framework of the Bean critical-state model, that the radial temperature gradient appearing in tablets or annuli during a pulsed field magnetization can explain the negative magnetic relaxation in the superconductor.*

Advances in the synthesis of large melt grown crystals of high-temperature superconductors (HTS) based on Y (Re) BaCuO [1] led to extensive investigations of their superconducting properties with a view of practical applications. One of these applications is creating magnetic systems for compact EPR and NMR spectrometers. The superconductors are magnetized either in a static magnetic field (FC) which is switched off afterwards, or using a pulse method. The pulsed field magnetization (PFM) method is the preferred one because of low power consuming and simplicity. The main disadvantage of HTS materials is a reduction of the critical current $J_c$ with time, which is caused by magnetic flux creep. Therefore, the trapped magnetic moment decreases with time, regardless of the magnitude of the magnetization, when superconductors are slowly (isothermally) magnetized. Generally this process is linear in a logarithmic time scale [2]. On the other hand, after a short PFM of HTS one can observe not only the decrease of the magnetic moment, but also its growth [3,4]. The growth of the trapped magnetic field is observed in tablets, as well as in annuli if they are not completely magnetized by a pulsed magnetic field. This effect can be called as a negative relaxation according to the accepted definition of the relaxation rate [2]. The negative relaxation is observed when the trapped field is less, by 10-20%, than the maximal value $B_{max}$.

Fig. 1 shows the experimental dependence of the normalized trapped field $B(t) / B_{max}$ in the gap of 2 mm between two superconducting annuli (manufactured by the method described in [5]) after the action of magnetizing pulses of 10 ms duration. The pulse amplitude was gradually increased during the multi-pulsed field magnetization (multi-PFM). The trapped field is normalized to the maximal attainable value of $B_{max} \approx$ 0,7T. At the initial stage of magnetization, when $B / B_{max} \ll 0.75$ (see Fig. 1C), we observed a jump in the trapped field just after a field pulse, followed by a continuous growth of B with the time. Fig. 1B shows the trapped field data after pulses for intermediate magnetization. In this case the rate of the magnetization growth is essentially reduced and after the jump the field changes only slightly with the time. In large fields (Fig. 1A) near the maximum magnetization ($B / B_{max} > 0.9$), the field B decreases with time. Note that reducing of the trapped field is typical for superconductors (a positive relaxation of the magnetization), in contrast to the negative relaxation observed at the initial part of the multi-PFM.

It is shown below that the negative relaxation is a result of a temperature gradient in the superconductor. To simplify the calculations, we consider the magnetization of a thick-walled superconducting cylinder with inner radius $R_1$ and outer radius $R_2$. For the calculation of currents and fields we use the Bean critical-state model (see [6]), which assumes that the critical current depends on temperature, but not on the magnetic field. The Bean model describes well the isothermal (slow) magnetization process.

On the other hand, when the pulse of the magnetization field is shorter, ≤ 10 ms, the thermal diffusion time is much longer than the field pulse length. In this case the magnetization can be considered as an adiabatic process [7], and the instantaneous temperature near the outer edge of the cylinder is higher than in the inner one. This is crucial in the calculation of the fields and currents distribution in the superconductor.

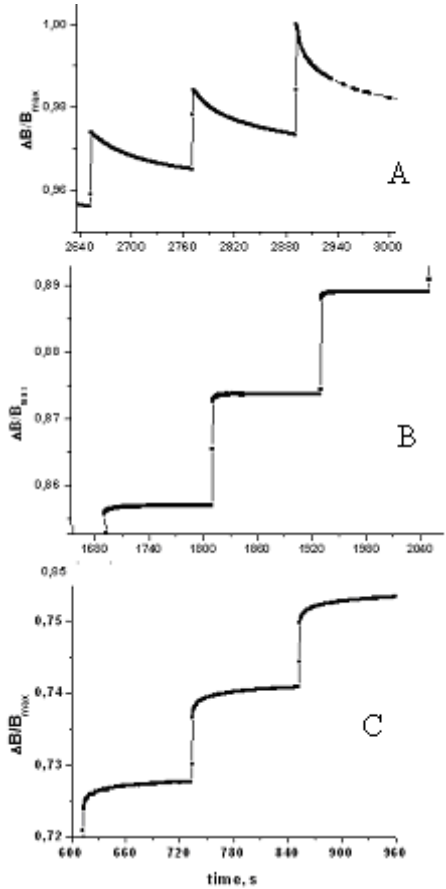

Fig. 1. Time evolution of the normalized magnetization B (t) / $B_{max}$ in the gap of 2 mm between two superconducting annuli after the pulse magnetization.

It is difficult to calculate exactly the warming of the superconductor during pulsed magnetization, since in this case the current-voltage characteristic is unknown. As a first approximation, we assume that immediately after the pulse the local temperature depends linearly on the cylinder's radius. Thus, T increases linearly from the inner edge of superconducting cylinder to the outer one, as it turns out in many calculations [8,9]. Assuming that the experiments are carried out at liquid nitrogen temperature, the radial heating $\delta T$ can be written as

$$T = 78 + \delta T$$
$$\delta T(r) = \Delta \cdot (r-R_1)/(R_2-R_1) \quad (1),$$

where $\Delta$ is the radial gradient of temperature or overheating of the outer wall toward the inside one.

It is usually taken as granted that the critical current decreases linearly with increasing temperature, $J_c = J_0 (1-T/T_c)$. In this case, the current density that determines the trapped flux is not constant, as in the simple Bean model, but decreases linearly with increasing radius [6]. Therefore, the critical current density in the inner region is higher than at the outer edge near $R_2$ that can be written as:

$$J_c = J_0 (1-(78+\delta T(r))/T_c) \quad (2)$$

It is known that the trapped field reaches its maximum value in the body of a superconductor if it is not completely magnetized [6]. In this case the current density distribution is given by

$$J_c = \begin{cases} -J_c(\delta T(r)), & R_1 < r < R_m \\ +J_c(\delta T(r)), & R_m < r < R_2 \end{cases} \quad (3),$$

which corresponds to the maximum trapped field at the distance $R_m$ in the cylinder's body.

The current distribution (3) is schematically depicted in Fig. 2.

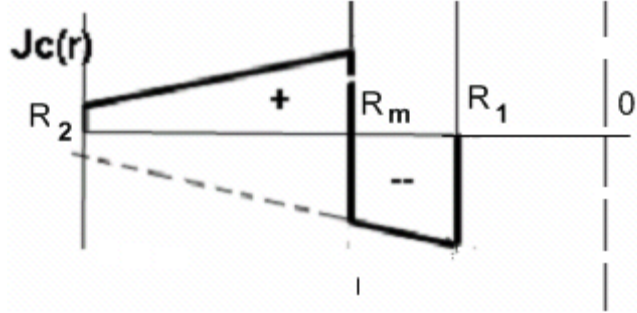

Fig. 2. Schematic distribution of critical currents in a superconducting hollow cylinder in the presence of temperature gradient.

The minus sign in Fig. 2 and Eq. (3) indicates the diamagnetic currents that occur on the rising edge of the pulse magnetization, and (+) corresponds to the currents which appear at the recession of the external field pulse. The currents change its direction at the point of current discontinuity $r = R_m$ where the trapped magnetic field has its maximum.

Superconducting currents create a trapped field in the cylinder hole as given by

$$h_0 = \int_{R_1}^{R_2} Jc(r, \delta T) dr \quad (4).$$

At the integration one should take into account the condition (3), with the change of sign of the current.

Due to a flux creep, the current density decreases linearly on the logarithmic time scale [2],

$$J = J_c(1 - S \cdot \ln(t/t_0)). \quad (5)$$

Here S is the so-called logarithmic relaxation rate defined by the derivative of $S = -d \ln (J) / d\ln (t)$. After cooling at the equilibrium temperature (T = 78 K) the relaxation rate of the superconducting currents depends on the distance r, since these currents had appeared at different temperatures. It is known that the current relaxation rate exponentially decreases if the relaxing superconductor is cooled at $\Delta T$ [2, 10]. For example, S reduces by 6 times after lowering the temperature from 78 K on $\Delta T = 2$ K and S reduces by 200 times when the temperature is decreased on $\Delta T = 4K$, respectively [11]. This corresponds to an exponential decrease of the rate of $S = S_{78} \cdot e^{-1,2\delta T}$, where $\delta T$ is given in degrees K.

In view of the pulsed heating of the superconductor (1), the relaxation rate after cooling decreases from the inner to the outer edge, and the distribution of S can be written as

$$S = S_0 \cdot e^{-1.2\delta T(r)} \quad (6)$$

Here So is the relaxation rate at T = 78 K, and $\delta T$ is measured in K. Taking into account the attenuation of the superconducting currents with relaxation rate given by (6), the time dependence of the trapped magnetic field can be written as

$$h(t) = h_o - \int Jc(r,T) \, S_0 \cdot e^{-1,2\delta T(r)} dr \cdot \ln(t/t_o). \quad (7)$$

By analogy with the superconducting currents we determine the field relaxation rate in the same form as (5). Accordingly, from (4) and (7) we obtain an expression for the relative relaxation rate of the trapped field in a hollow cylinder

$$\frac{S}{So} = \frac{ho - h}{ho \cdot \ln(t/to)} = \frac{\int_{R1}^{R2} Jc(r,T) e^{-\delta T} dr}{\int_{R1}^{R2} Jc(r,T) dr} \quad (8)$$

In order to compare our model with experimental data we calculated the dependence of $S(R_m) / S_{78}$ for a hollow cylinder, in which $R_1 = 0.8$ cm, and $R_2 = 1.8$ cm. Annuli with such dimensions have been studied in [4]. Fig. 1 illustrates some of these data. The calculation results $S(R_m) / S_o$ depending on the position $R_m$ of the maximal trapped field are shown in Fig. 3. During the multi-PFM the maximal field in the superconductor shifts from the outer to the inner wall. The field in the hole of the cylinder occurs when $R_m \leq (R_1 + R_2) / 2$ [6], so in Fig. 3 the range of $R_m$ extends from the inner radius $R_1$ up to a half of the wall thickness. The lower curve corresponds to the overheating of the outer wall on $\Delta = 10$ K and the upper on $\Delta = 1$ K. For $\Delta = 0$, the relaxation rate, obviously, is constant (independent of $R_m$), and the $S/S_{78} = 1$. In contrast, for large $\Delta$ (strong overheating) S rapidly decreases with increasing $R_m$ and changes sign from positive to negative even at a short distance $R_m$ from $R_1$ (lower curves in Fig. 3). In this case, the higher value of the superheat $\Delta$ results in the broader region of the negative relaxation.

From geometrical considerations of the field distribution within the Bean model [3,6] it is easy to find the relation between the field in the hole of the cylinder B to its maximal value $B_{max}$, and the position $R_m$ of the largest field inside the body of the superconducting cilinder

$$B/B_{max} = 1 - 2(R_m - R_1)/(R_2 - R_1). \quad (9)$$

As seen in Fig. 1 the relaxation rate in the experiment changes sign at $B/B_m = 0.88$. From (9) we easily find that this corresponds to $R_m \approx 0.86$. The calculation with formula (8) shows (see Fig. 3) that for $R_m = 0.86$ the relaxation rate S is zero in case of overheating $\Delta \approx 7K$. According to the published data a typical overheating for a pulse magnetization consists of a few degrees [12,13]. Due to the lack of accurate data on the values of overheating it is difficult to talk about the quantitative aspect of the described model, but the qualitative agreement is evident between the presented rough approximations and the experimental data.

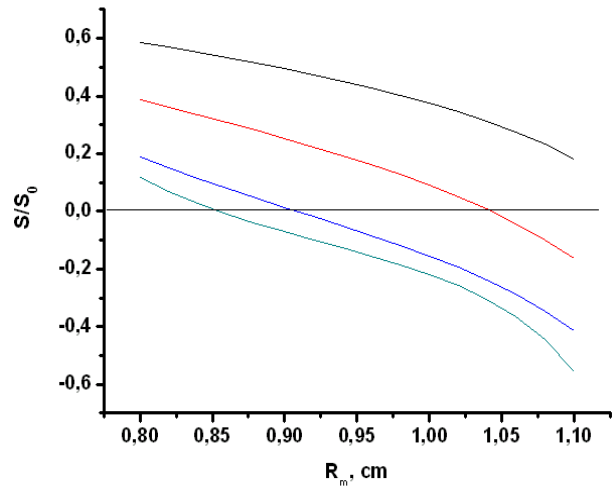

Fig. 3. The relative relaxation rate of the field in the bore of the hollow cylinder, depending on the position of $R_m$. The curves correspond from top to bottom $\Delta$ = 1K, 2K, 5K and 10K.

In conclusion, the temperature gradient appearing in PFM strongly effects the current relaxation in the superconductor. The currents near the outer surface diminish slower than the internal currents, which vary in the opposite direction. This can lead to the negative relaxation of superconducting currents, that is to an increasing of the magnetic moment of the superconductor as a function of time. Negative relaxation can be used to improve the time stability of the magnetic field in systems on basis melt grown HTS.

The author is grateful A.A.Kartamyshev and V.S.Korotkov for assistance in the measurements.

**Literature**

1. Yoo S I, Sakai N, Takaichi H, Higuchi T, Murakami M, Appl. Phys. Lett. **65** (1994) 633–5
2. Y.Yeshurun, A.P.Malozemoff, A.Shaulov, Rev. Mod. Phys., 68, N3 (1996) 911
3. E.P. Krasnoperov, A.A. Kartamyshev, Y.D.Kuroedov, O.L. Polushchenko, N.A. Nizelskij, Physica C 469 (2009) 805–809
4. E.P. Krasnoperov, A.A. Kartamyshev, D.I. Puzanov, O.L. Polushchenko, N.A. Nizelskij, J Supercond Nov Magn (2010) 23: 1499–1501
5. N.A. Nizhelskiy, O.L. Poluschenko, V.A. Matveev, Supercond. Sci. Technol., 20 (2007) 81
6. M.N.Wilson "Superconducting Magnets", Clarendon Press.Oxford 1983
7. A.Vl.Gurevich, R.G.Mints and A.L.Rakhmanov, "Physics of Composite Superconductors", Begell House, N Y, 1997
8. Masanori Tsuchimoto and Kenta Morikawa, IEEE Transaction on applied superconductivity, v. 9, N. 1, Marh 1999
9. S. Bræck, D. V. Shantsev, T. H. Johansen, Y.M.Galperina, J Appl. Phys., v. 92, N 10
10. Weinstein R, Liu J, Ren Y, Sawh R-P, Parks D, Foster C and Obot V, *Proc. 10th Anniversary HTS Workshop on Physics, Materials and Applications (Houston, TX, March,1996)* (Singapore: World Scientific) p. 625
11. Krabbes G., Fuchs G., Canders W-R., May H., Palka R. "High Temperature Superconductor Bulk Materials", WILEY-VCH, 2006
12. Y Yanagi, Y Itoh, M Yoshikawa, T Oka, H Ikuta, U Mizutani Supercond. Sci. Technol. **18** (2005) 839–849
13. H. Fujishiro, T. Naito, D. Furuta, K. Kakehata, Physica C 470 (2010) 1181–1184